\documentclass[twocolumn,prb,showpacs,preprintnumbers,amsmath,amssymb]{revtex4}

\usepackage{graphicx}
\usepackage{dcolumn}
\usepackage{bm}

\begin{document}


\title{Diamagnetism around the Meissner transition in a homogeneous cuprate single crystal}

\author{Jes\'us Mosqueira}
\author{Luc\'ia Cabo}%
\author{F\'elix Vidal}
\affiliation{LBTS, Departamento de F\'isica da Materia Condensada, Universidade de Santiago de Compostela, E-15782 Santiago de Compostela, Spain}

\date{\today}

\begin{abstract}
The in-plane diamagnetism around the Meissner transition was measured in a Tl$_2$Ba$_2$Ca$_2$Cu$_3$O$_{10}$ single crystal of high chemical and structural quality, which minimizes the inhomogeneity and disorder rounding effects on the magnetization. When analyzed quantitatively and consistently above and below the transition in terms of the Ginzburg-Landau (GL) approach with fluctuations of Cooper pairs and vortices, these data provide a further confirmation that the observed Meissner transition is a conventional GL superconducting transition in a homogeneous layered superconductor.
\end{abstract}

\maketitle

\section{Introduction}

An important and still open issue, at present under considerable discussion, of the high-$T_C$ cuprate superconductors (HTSC) is the role that play in their superconducting phase transition the chemical and electronic disorder and inhomogeneities,\cite{uno} and the loss of long-range phase coherence due to vortex fluctuations.\cite{dos} The ongoing debate, whose elucidation may be an appreciable step toward the understanding of the pairing mechanism in HTSC, is particularly well illustrated by the \textit{precursor} diamagnetism observed above the Meissner transition. In optimally doped HTSC, this effect was first studied quantitatively by Lee, Klemm and Johnston \cite{tres} and attributed, as in the case of the conventional (BCS) low-$T_C$ superconductors,\cite{cuatro} to the presence of evanescent Cooper pairs created by the unavoidable thermal agitation. This last conclusion was supported by the fact that the measured diamagnetism could be accounted for in terms of the Ginzburg-Landau approach for multilayered superconductors with Gaussian fluctuations of the superconducting order parameter (GGL approach). Since then, the applicability of this \textit{conventional} phenomenological GGL scenario, which was more recently extended empirically to high reduced temperatures by the introduction of a \textit{total energy} cutoff,\cite{cinco} has been confirmed by different groups in many HTSC,\cite{seis,siete} very in particular in underdoped compounds up to relatively high reduced field amplitudes, of the order of 0.2.\cite{ocho,ochoa} Further support was earlier provided by the consistency with other intrinsic rounding effects observed above $T_C$, especially with the paraconductivity,\cite{siete,Welp} and also with the precursor diamagnetism in dirty low-$T_C$ superconductors.\cite{nueve}

In the last few years, however, the results summarized above are being strongly questioned, the claimed disagreements being with both the experimental and the theoretical aspects.\cite{diez,once,doce,trece,quince,dieciseis,diecisiete,dieciocho,veinte} Among the first ones, is the anomalous precursor diamagnetism observed in various HTSC with different dopings, including the optimal one.\cite{diez,once,veinte} These anomalies are being proposed as a further confirmation of \textit{unconventional} (non-GGL) scenarios, including those based on the presence of intrinsic $T_C$-inhomogeneities,\cite{doce,quince} on the breaking of the long-range phase coherence by vortex fluctuations,\cite{trece,dieciseis,diecisiete} or on a 3D Bose-Einstein condensation.\cite{dieciocho} However, it has been shown recently that at least some of these anomalies may be easily explained in terms of $T_C$ inhomogeneities associated with the presence of chemical inhomogeneities in the samples.\cite{diecinueve} In fact, intrinsic-like chemical inhomogeneities and disorder, and then the corresponding $T_C$ inhomogeneities, may be always present in some cuprates (in particular, in underdoped compounds), due to their non-stoichiometric nature.\cite{uno} Nevertheless, even when the measured precursor diamagnetism agrees with the GGL predictions, some authors contest the adequacy of these analysis invoking the apparently anomalous magnetization behavior below $T_C$ also in optimally doped cuprates.\cite{veinte,comment}
      
The challenge is, therefore, first to disentangle the diamagnetism due to superconducting fluctuations from the magnetization roundings associated with, intrinsic or not, chemical disorder and inhomogeneities, and then to attempt to analyze these data, consistently above and below $T_C$, in terms of vortex and Cooper pairs fluctuations. This is the purpose of this paper, where we will first summarize the magnetization measurements in a Tl$_2$Ba$_2$Ca$_2$Cu$_3$O$_{10}$ (Tl-2223) single crystal with very high chemical and structural quality. This compound is stoichiometric, which minimizes the randomness of dopant atom distribution and then the corresponding disorder and inhomogeneities.\cite{uno} These data will be then used to check if the existing GL approaches with fluctuations of vortices and of Cooper pairs \cite{tres,siete,ocho,ochoa,veintiuno,veintidos,veintitres} explain, at quantitatively and consistently above and below $T_C$, the diamagnetism measured in homogeneous cuprate superconductors.

\section{Experimental details and results}

The Tl-2223 sample used in this work, a $1.1\times0.75\times0.226$ mm$^3$ single crystal, was chosen by its sharp low-field Meissner transition from several high-quality single crystals of this nominal composition prepared by A. Maignan and coworkers in Caen. Details of the growth procedure of these crystals and of their structural characterization may be found in Ref. \onlinecite{veinticuatro}. The temperature and magnetic field dependence of the magnetization, $M (T, H)$, around $T_C(H)$ were measured with a superconducting-quantum-interference-device (SQUID) magnetometer (Quantum Design). A crucial check of the quality of our crystal is provided by the temperature dependence of its low-field (1 mT) magnetic susceptibility measured under field-cooled (FC) and zero field-cooled (ZFC) conditions (inset of Fig.~1). Under these very low fields, the superconducting fluctuation effects are expected to be negligible \cite{tres,cuatro,cinco,seis,siete,veintiuno,veintidos,veintitres} and, therefore, the observed susceptibility roundings above $T_C$ will be mainly due to inhomogeneities and disorder. As it may be appreciated, the Meissner transition is very sharp, the transition temperature width, defined as usually,\cite{diecinueve} being around $\sim 1$ K, and the zero-field critical temperature $T_{C0} = 122.0$ K. These data also show an exceptionally high Meissner fraction, $|\chi^{FC}(T\to0)|\approx 0.8$. 

\begin{figure}[b]
\includegraphics[scale=0.5]{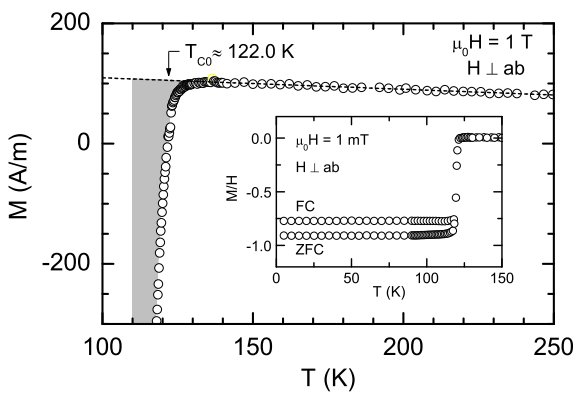}
\caption{$T$ dependence of the as-measured magnetization, for $\mu_0H=1$ T perpendicular to the \textit{ab} planes. The line is the background contribution, obtained by fitting a Curie-like function in the region 150-250 K. The shadow area represents the effect of the superconducting fluctuations. Inset: $T$ dependence of the low-field magnetic susceptibility, already corrected for demagnetizing effects, measured under FC and ZFC conditions.}
\end{figure}

\begin{figure}[t]
\includegraphics[scale=0.5]{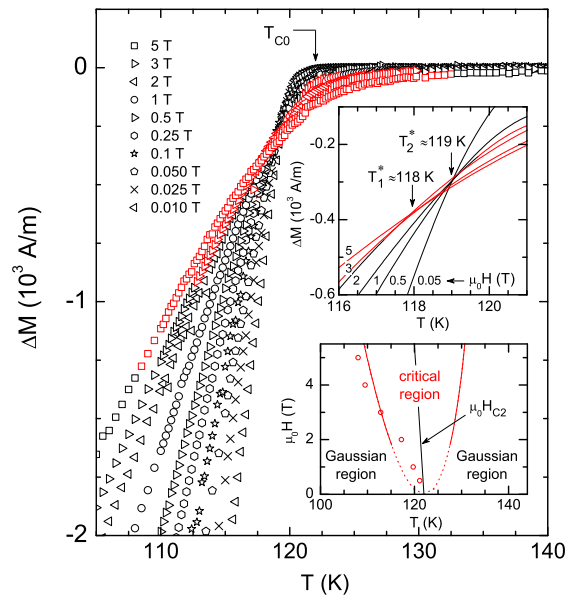}
\caption{(Color online) $T$ dependence of the excess magnetization around $T_C$. Data in the critical region around $H_{C2}(T)$ are indicated in red. Upper inset: Detail of the crossing point which reveals its splitting: the $\Delta M(T)$ curves into the critical region cross at $T_1^*$, while the ones outside cross at $T_2^*$. Lower inset: $H-T$ phase diagram for the fluctuations. The red curve is the critical region boundary as estimated from Eq.~(2). The data points represent the experimental window for the scaling approach in the critical region. See the main text for details.}
\end{figure}

The \textit{in-plane} diamagnetism (for $H$ applied perpendicularly to the crystallographic \textit{ab} planes) may be parametrized through the so-called \textit{excess magnetization}, $\Delta M(T,H)$, which as usual is defined as the difference between the as-measured magnetization and the normal-state or background contribution, $M_B(T,H)$. This last was obtained by fitting to the $M(T,H)$ curves well above $T_C$ (between $\sim$150 K and $\sim$250 K) a Curie-like function (see Fig.~1 for an example corresponding to $\mu_0H=1$ T). An overview of the measured $\Delta M(T)_H$ curves is presented in Fig.~2. These data extend to both sides of $T_C$ and cover up to 2.5 orders of magnitude in $H$, which allows a thorough check of the GL approaches in the different regions, where the influence of fluctuating Cooper pairs and vortices  are expected to be different.\cite{tres,ocho,ochoa,veintiuno,veintidos,veintitres} As shown in this figure, the $\Delta M(T)_H$ curves cross at a temperature few degrees below $T_{C0}$,\cite{Kes} which is a signature of the thermal fluctuations in highly anisotropic superconductors. When expanding these data (upper inset of Fig. 2), the crossing point spans a temperature interval between $T_1^*\approx 118$ K for high fields and $T_2^*\approx 119$ K for low fields. A similar splitting has been also observed in other highly anisotropic superconductors \cite{Naughton} and will be analyzed below in detail. 

\section{Data analysis}

Above but not too close to $T_C$, on the grounds of the GGL scenario in the 2D limit, well adapted to this highly anisotropic superconductor, $\Delta M$ is given by \cite{cinco,ochoa}
\begin{eqnarray}
\Delta M=-f\frac{k_BTN}{\phi_0s}\left[-\frac{\varepsilon^c}{2h}\psi\left(\frac{h+\varepsilon^c}{2h}\right)-\ln\Gamma\left(\frac{h+\varepsilon}{2h}\right)\right.\nonumber \\
+\left.\ln\Gamma\left(\frac{h+\varepsilon^c}{2h}\right)+\frac{\varepsilon}{2h}\psi\left(\frac{h+\varepsilon}{2h}\right)+\frac{\varepsilon^c-\varepsilon}{2h}\right].
\label{Prange}
\end{eqnarray}
Here $\Gamma$ and $\psi$ are the gamma and digamma functions, $h\equiv H/H_{C2}(0)$ the reduced magnetic field, $H_{C2}(0)$ the upper critical field extrapolated to $T = 0$ K, $\varepsilon\equiv\ln(T/T_{C0})$ the reduced temperature, $N=3$ the number of superconducting CuO$_2$ layers in their periodicity length ($s=c/2=17.9$ \r{A}), $f\approx0.8$ the effective superconducting volume fraction (approximated as the Meissner fraction \cite{volume_fraction}), $k_B$ the Boltzmann constant, $\phi_0$ the flux quantum, and $\varepsilon^c\approx 0.55$ the total-energy cutoff constant.\cite{cinco} The solid lines in Fig.~3(a) are fits of Eq.~(1) in the $\varepsilon$-region $2\times10^{-2}\stackrel{<}{_\sim}\varepsilon\stackrel{<}{_\sim}2\times10^{-1}$ to some of the data above $T_C$ of Fig.~2, with $H_{C2}(0)$ as the only free parameter. These few examples show an excellent agreement down to a few degrees above $T_{C0}$, where the Gaussian approximation is no longer valid (see below), and it leads to $\mu_0H_{C2}(0)=340$ T, a value that is going to be used in the remaining analysis above and below $T_C$. As illustrated by the examples shown in the inset of Fig.~3(a), the agreement of Eq.~(1) with the $H$ dependence of $\Delta M$ measured above $T_C$ is also excellent.

\begin{figure}[t]
\includegraphics[scale=0.55]{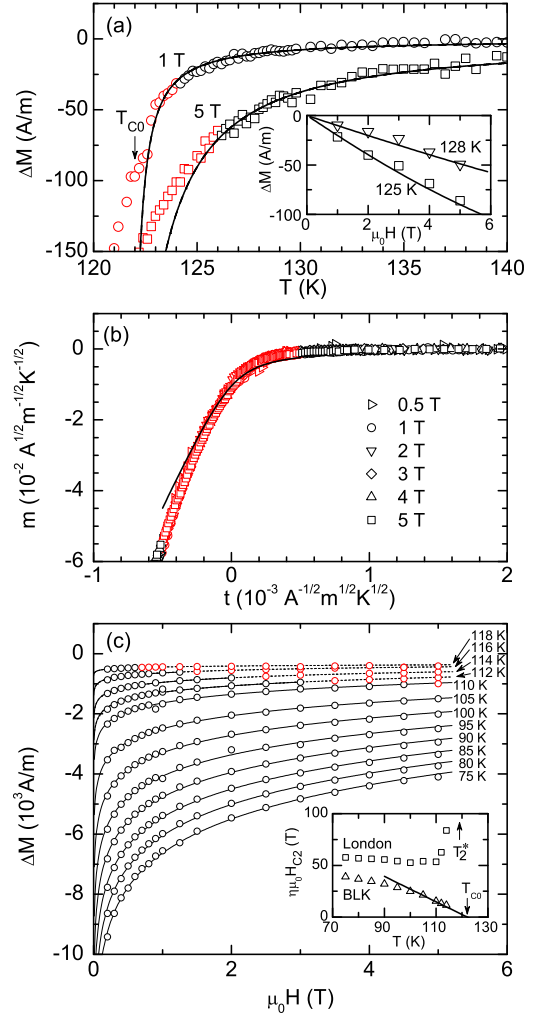}
\caption{(Color online) a) Two examples of the excess magnetization vs. $T$ and vs. $H$ (inset) in the Gaussian region above $T_C(H)$. The lines are the GGL prediction [Eq.~(\ref{Prange})]. b) Scaling of the $m$ vs $t$ curves in the critical region around $H_{C2}(T)$. The line is the GL-LLL scaling function [Eq.~(\ref{scalingfunction})]. c) $\Delta M$ vs $H$ in the Gaussian region below $H_{C2}(T)$. The lines are fits to the BLK theory [Eq.~(6)]. The resulting $\eta H_{C2}(T)$ is presented in the inset and close to $T_{C0}$ it has the temperature behavior and amplitude predicted by the GL approach. In contrast, if the conventional London term alone is used, $\eta H_{C2}(T)$ diverges at $T^*_2$. In all figures, data points in red are in the critical region. }
\end{figure}

Near $T_C(H)$ the fluctuations increase so much that the Gaussian approximation breaks down. In a magnetic field sufficiently strong that the fluctuations' spectrum is restricted to the lowest Landau-level (LLL), this critical region for two-dimensional systems is bounded by \cite{Ikeda}
\begin{equation}
\frac {\left|T-T_C(H)\right|}{T_{C0}}\stackrel{<}{_\sim}\sqrt{\frac{2k_B}{\Delta c\xi_{ab}^2(0)s}h},
\label{criterio}
\end{equation}
where $\Delta c$ is the specific heat jump at $T_{C0}$, and $\xi_{ab}(0)$ the in-plane superconducting coherence length amplitude. By using\cite{deltac} $\Delta c=3\times10^5$ J/Km$^3$  and $\xi_{ab}(0)=9.8$ \r{A} (which corresponds to the $H_{C2}(0)$ value obtained above in the GGL region above $T_C$), the limits of this \textit{critical} region are represented in the lower inset of Fig.~2. In this LLL regime, the GL theory predicts that for two-dimensional systems $\Delta M$ follows a scaling behavior in the variables \cite{veintiuno}
\begin{equation}
m\equiv \frac{\Delta M}{\sqrt{HT}},\;\;\;t\equiv \frac{T-T_C(H)}{\sqrt{HT}},
\label{scaling}
\end{equation}
the scaling function being calculated by Te\u{s}anovi\'{c} and coworkers as \cite{veintitres}
\begin{equation}
m=f\frac{A}{H_{C2}^{'}}\frac{k_B}{\phi_0s}\left(At-\sqrt{A^2t^2+2}\right),
\label{scalingfunction}
\end{equation}
where $A\equiv[H_{C2}^{'}T^*_1/2(T_{C0}-T^*_1)]^{1/2}$, $H_{C2}^{'}\equiv H_{C2}(0)/T_{C0}$, and $T^*_1$ is close to the limit of the critical region below $T_{C0}$ when $H=0$. Equation (4) predicts a crossing of the $M(T)_H$ curves at $T^*_1$ and $\Delta M_1^*$, related through
\begin{equation}
\Delta M^*_1=-f\frac{k_BT^*_1}{\phi_0s},
\label{crucealto}
\end{equation}
which allows a first direct comparison with the results of Fig.~2. Note first that the high-field crossing point observed at $T_1^*$ (upper inset in Fig.~2) falls into the critical region delimited by Eq.~(\ref{criterio}) and, therefore, should be described by Eq.~(\ref{crucealto}). By using $T_1^*=118$ K, this equation leads to $\Delta M^*_1\approx-340$ A/m, which is in good agreement with the experimental value ($-375$ A/m), taking into account the experimental uncertainties in $f$ and in the normal-state $M_B(T)$ contribution. In Fig.~3(b) we represent the $\Delta M(T,H)$ data in the critical region in terms of the $t$ and $m$ variables given by Eq.~(3), calculated by using $T_C(H)=T_{C0}[1-H/H_{C2}(0)]$ and $\mu_0H_{C2}(0)=340$ T, the value obtained before in the GGL region above $T_C$. As can be seen, the scaling of the $m(t)$ curves is excellent and the scaling function, evaluated from Eq.~(4) by using again the same parameters, is also in agreement with the data at a quantitative level. A good agreement with the GL-LLL scaling was also observed by various groups in different optimally doped and intrinsically underdoped cuprates.\cite{Welp,Sok,MunKim,Lascialfari}

For temperatures well below $T_C(H)$, outside the critical region, the fluctuation diamagnetism will be dominated by fluctuations of the two-dimensional vortex (\textit{pancakes}) positions, which in terms of the GL approach may be seen as fluctuations of the order parameter phase. This contribution has been calculated by Bulaevskii, Ledvig and Kogan (BLK approach) as,\cite{veintidos}
\begin{eqnarray}
\Delta M(T,H)=-f\frac{\phi_0}{8\pi\mu_0\lambda^2_{ab}(T)}{\rm ln}\left(\frac{\eta H_{C2}(T)}{H}\right)+\nonumber\\
+f\frac{k_BT}{\phi_0s}{\rm ln}\left(\frac{8\pi\mu_0k_BT\lambda_{ab}^2(T)}{\alpha s\phi_0^2\sqrt{e}}\frac{H_{C2}(T)}{H}\right).
\label{BLK}
\end{eqnarray}
The first term on the right is the conventional London magnetization, whereas the second one is associated with fluctuations. In this equation, $\lambda_{ab}$ is the magnetic penetration length in the $ab$ planes, $\eta$ and $\alpha$ are constants of the order of the unity related to the vortex structure, and $\sqrt{e}\approx1.649$ . This equation also predicts the crossing of the $\Delta M(T)_H$ curves at a temperature $T_2^*$ a few degrees below $T_{C0}$, the corresponding excess magnetization being
\begin{equation}
\Delta M^*_2=-f\frac{k_BT^*_2}{\phi_0s}\ln{(\eta\alpha\sqrt{e})},
\end{equation}
which differs from Eq.~(5) for the crossing point in the critical region, by a constant of the order of the unity. By using in Eq.~(7) the experimental $T_2^*$ and $\Delta M_2^*$ values (upper inset of Fig.~2), which are well outside the critical region, one finds $\ln(\eta\alpha\sqrt{e})\approx0.8$, in agreement with the BLK approach. 

A thorough comparison of Eq.~(\ref{BLK}) with the experimental data is presented in Fig.~3(c). For each isotherm, the only free parameters are $\eta H_{C2}$ and $\lambda_{ab}$. Clearly, the fit quality is excellent for isotherms up to $\sim T_2^*$. As shown in the inset of Fig.~3(c), close to $T_{C0}$ the resulting $\eta H_{C2}$ follows a linear temperature dependence and tends to zero at $T_{C0}$, in full agreement with the conventional (GL) behavior for $H_{C2}(T)$. Moreover, by using the previous $\mu_0H_{C2}(0)=340$ T, one obtains, as expected, $\eta\approx0.4$.  It is worth noting that the London theory alone (without the term due to vortex fluctuations) fits the data of Fig.~3(c) as well (it follows the same $H$ dependence). However, as earlier stressed by Kogan and coworkers,\cite{Kogan} it would lead to an anomalous $H_{C2}(T)$ close to $T_{C0}$, due to the presence of the crossing point. This seemingly anomalous $H_{C2}(T)$ behavior has been recently presented as the hallmark of unconventional behavior for $\Delta M$.\cite{veinte}

\section{Conclusions}

We have presented detailled measurements of the diamagnetism around the Meissner transition in a high quality Tl-2223 single crystal. These results were analyzed in terms of the Ginzburg-Landau approach with superconducting fluctuations in the three different fluctuation regions: the Gaussian-Ginzburg-Landau regions above and below $T_C$ and the critical region closer to $T_C(H)$. These analysis extend previous comparisons, up to now performed consistently only in one or two of these fluctuation regions.\cite{tres,cuatro,cinco,seis,siete,ocho,ochoa,Welp} They provide, therefore, an important test of consistency for the applicability of the GL approaches to describe the Meissner transition in an optimally doped cuprate. When combined with previous measurements of different observables around $T_C$,\cite{tres,cuatro,cinco,seis,siete,ocho,ochoa,Welp} in particular of $\Delta M$ in underdoped cuprates,\cite{ocho,ochoa,diecinueve,lasco,Sok,MunKim,Lascialfari} our present results provide a strong experimental confirmation that the Meissner transition in all cuprates is, independently of the doping, a conventional GL transition, although in many cases entangled with, intrinsic or not, disorder and inhomogeneities. This work may then help to reach a general consensus on the phenomenological descriptions of the superconducting transition in cuprates, which in turn will provide a constraint for the explanations of the critical behavior of any observable around $T_C$ \cite{nota} and, more important, for any proposal of pairing mechanism in these materials.

\section{Acknowledgements}

We thank A. Maignan and A. Wahl for providing us the Tl-2223 crystals, the Spanish Ministerio de \mbox{Educaci\'on} y Ciencia (Grant No. MAT2004-04364) and the Xunta de Galicia (Grant No. PGIDIT04TMT206002PR). L.C. acknowledges support through a FPU grant.


\begin{references}


\bibitem{uno}	See, e.g., E. Dagotto, Science {\bf 309}, 257 (2005), and references therein; see also, K. McElroy, Jinho Lee, J.A. Slezak, D.-H. Lee, H. Eisaki, S. Uchida and J.C. Davis, \textit{ibid.} {\bf 309}, 1048 (2005).

\bibitem{dos}	See, e.g., J. Orenstein and A.J.  Millis, Science {\bf 288}, 468, (2000); see also,  P.A. Lee, N. Nagaosa and X. G. Wen,  Rev. Mod. Phys. {\bf 78}, 17 (2006), and references therein.

\bibitem{tres}W.C. Lee, R.A. Klemm and D.C. Johnston, Phys. Rev. Lett. \textbf{63}, 1012 (1989).

\bibitem{cuatro}See, e.g., M. Tinkham in \textit{Introduction to Superconductivity} (McGraw-Hill, New York) 1996, Chapt. 8.

\bibitem{cinco}J. Mosqueira, C. Carballeira, M.V. Ramallo, C. Torr\'on, J.A. Veira and F. Vidal, Europhys. Lett. {\bf 53}, 632 (2001); F. Vidal, C. Carballeira, S.R. Curr\'as, J. Mosqueira, M.V. Ramallo, J.A. Veira and J. Vi\~na, \textit{ibid.} {\bf 59}, 754 (2002).

\bibitem{seis}See e.g., Q. Li, in {\it Physical Properties of High Temperature Superconductors V}, edited by D.M. Ginsberg (World Scientific, 1996), Chap. 4; see also, A. A. Varlamov, G. Balestrino, E. Milani, and D. V. Livanov, Adv. Phys. \textbf{48}(6), 655 (1999).

\bibitem{siete}See e.g., F. Vidal and M.V. Ramallo in \textit{The Gap Symmetry and Fluctuations in High-Tc Superconductors}, Vol. 371 of NATO Advanced Study Institute, Series B: Physics, edited by J. Bok \textit{et al.}, Plenum, New York, 1998, p.443.

\bibitem{ocho}C. Carballeira, J. Mosqueira, A. Revcolevschi and F. Vidal, Phys. Rev. Lett. {\bf 84}, 3157 (2000).

\bibitem{ochoa}C. Carballeira, J. Mosqueira, A. Revcolevschi and F. Vidal, Physica C {\bf 384}, 185 (2003).

\bibitem{Welp}See, e.g., U. Welp, S. Fleshler, W.K. Kwok, R.A. Klemm, V.M. Vinokur, J. Downey, B. Veal, and G.W. Crabtree, Phys. Rev. Lett. \textbf{67}, 3180 (1991).

\bibitem{nueve}	J. Mosqueira, C. Carballeira and F. Vidal, Phys. Rev. Lett. {\bf 87}, 167009 (2001); F. Soto, H. Berger, L. Cabo, C. Carballeira, J. Mosqueira, D. Pavuna, and F. Vidal, Phys. Rev. B \textbf{75}, 094509 (2007).

\bibitem{diez}P. Carretta, A. Lascialfari, A. Rigamonti, A. Rosso, and A. Varlamov, Phys. Rev. B {\bf 61}, 12420 (2000); A. Lascialfari, A. Rigamonti, L. Romano, A.A. Varlamov, and I. Zucca, \textit{ibid.} {\bf 68}, 100505(R) (2003), and references therein.

\bibitem{once}L. Li, Y. Wang, M.J. Naughton, S. Ono, Y. Ando, and N.P. Ong, Europhys. Lett. \textbf{72}, 451 (2005).

\bibitem{veinte}Y. Wang, L. Li, M.J. Naughton, G.D. Gu, S. Uchida, and N.P. Ong, Phys. Rev. Lett. \textbf{95}, 247002 (2005). 

\bibitem{doce}V.Z. Kresin, Y.N. Ovchinikov and S.A. Wolf, Phys. Rep. \textbf{431}, 231 (2006), and references therein.

\bibitem{quince}J.L. Gonz\'alez and E.V.L. de Mello, Phys Rev. B \textbf{69}, 134510 (2004), and references therein.

\bibitem{trece}	A. Sewer and H. Beck, Phys. Rev. B \textbf{64}, 014510 (2001).

\bibitem{dieciseis}P.W. Anderson, Phys. Rev. Lett. \textbf{96}, 017001 (2006); cond-mat/0504453 (unpublished).

\bibitem{diecisiete}V. Oganesyan, D.A. Huse and S. L. Sondhi, Phys. Rev. B \textbf{73}, 094503 (2006). 

\bibitem{dieciocho}A.S. Alexandrov, Phys. Rev. Lett. \textbf{96}, 147003 (2006). 

\bibitem{diecinueve}L. Cabo, F. Soto, M. Ruibal, J. Mosqueira, and F. Vidal, Phys. Rev. B \textbf{73}, 184520 (2006). For a review on the rounding effects around the superconducting transition associated with $T_C$ inhomogeneities at long length scales see F. Vidal, J. A. Veira, J. Maza, J. Mosqueira, and C. Carballeira, \textit{Materials Science, Fundamental Properties and Future Electronic Applications of High-Tc Superconductors}, NATO ASI series, edited by S.L. Dreschler and T. Mishonov (Kluwer-Dordrech, Amsterdam) 2001, p. 289. A recent version of this paper may be seen in arxiv:cond-mat/0510467.


\bibitem{comment}L. Cabo, J. Mosqueira and F. Vidal, Phys. Rev. Lett. \textbf{98}, 119701 (2007); N.P. Ong, Y. Wang, L. Li, and M.J. Naughton, \textit{ibid.} \textbf{98}, 119702 (2007).

\bibitem{veintiuno}S. Ullah and A. Dorsey, Phys. Rev. Lett. \textbf{65}, 2066 (1990); Phys. Rev. B \textbf{44}, 262 (1991).

\bibitem{veintidos}L. N. Bulaevskii, M. Ledvig and V.G. Kogan, Phys. Rev. Lett. \textbf{68}, 3773 (1992).

\bibitem{veintitres}Z. Te\u{s}anovi\'c, L. Xing, L. Bulaevskii, Q. Li, and M. Suenaga, Phys. Rev. Lett. \textbf{69}, 3563 (1992).

\bibitem{veinticuatro}A. Maignan, C. Martin. V. Hardy, Ch. Simon, M. Hervieu, and B. Raveau, Physica C \textbf{219}, 407 (1994); some earlier magnetization measurements may be seen in F. Vidal et al., in {\it Superconducting and Related Oxides: Physics and Nanoengineering III}, edited by D. Pavuna and I. Bozovic, Proceedings of SPIE Volume 3481 (1998), p. 32.

\bibitem{Kes}P.H. Kes, C.J. van der Beek, M.P. Maley, M.E. McHenry, D.A. Huse, M.J.V. Menken, and A.A. Menovsky, Phys. Rev. Lett. {\bf 67}, 2383 (1991).

\bibitem{Naughton}M.J. Naughton, Phys. Rev. B {\bf 61}, 1605 (2000).

\bibitem{volume_fraction}J. Mosqueira, J.A. Camp\'a, A. Maignan, I. Rasines, A. Revcolevschi, C. Torr\'on, J.A. Veira, and F. Vidal, Europhys. Lett. {\bf 42}, 461 (1998).

\bibitem{Ikeda}R. Ikeda, T. Ohmi and T. Tsuneto, J. Phys. Soc. Jpn. {\bf 58}, 1377 (1989); \textit{ibid.} {\bf 59}, 1397 (1990); D.H. Kim, K.E. Gray, and M.D. Trochet, Phys. Rev. B {\bf 45}, 10801 (1992).

\bibitem{deltac}E. van Heumen, R. Lortz, A.B. Kuzmenko, F. Carbone, D. van der Marel, X. Zhao, G. Yu, Y. Cho, N. Barisic, M. Greven, C.C. Homes, and S.V. Dordevic, Phys. Rev. B \textbf{75}, 054522 (2007).

\bibitem{Sok}Junghyun Sok, Ming Xu, Wei Chen, B.J. Suh, J. Gohng, L.A. Schwartzkopf, B. Dabrovski, Phys. Rev. B {\bf 51}, 6035 (1995).

\bibitem{MunKim}Mun-Seog Kim, Wan-Seon Kim, Sung-Ik Lee, Seong-Cho Yu, Jin-Tae Kim, B. Dabrowski, J. Appl. Phys. {\bf 81}, 4231 (1997).

\bibitem{Lascialfari}A. Lascialfari, P. Tedesco and I. Zucca, Int. J. Mod. Phys. B {\bf 17}, 805 (2003).

\bibitem{Kogan}V.G. Kogan, M. Ledvij, A.Yu. Simonov, J.H. Cho, and D.C. Johnston, Phys. Rev. Lett. {\bf 70}, 1870 (1993).

\bibitem{lasco}J. Mosqueira, M.V. Ramallo, A. Revcolevschi, C. Torr\'on, and F. Vidal, Phys. Rev. B \textbf{59}, 4394 (1999).

\bibitem{nota}Just as an example, our present results do not support an anomalous (non-GL) superconducting transition in cuprates and, therefore, they favor the explanations of the high Nernst signal observed above $T_C$ in HTSC on the grounds of the GGL approach or as a normal state effect. See e.g., I. Ussishkin, S.L. Sondhi, and D. Huse, Phys. Rev. Lett. \textbf{89}, 287001 (2002); A.S. Alexandrov and V.N. Zavaritsky, \textit{ibid.} \textbf{93}, 217002 (2004); \textit{ibid.} \textbf{95}, 259704 (2005).



\end{references}
\end{document}